\documentclass[prl,aps,twocolumn,superscriptaddress]{revtex4-1}
\usepackage{graphicx,color}
\usepackage{amsthm}
\usepackage{amsfonts}
\usepackage{algorithmic}
\usepackage{enumerate}
\usepackage{latexsym}
\usepackage{amsmath}
\usepackage{amssymb}
\usepackage[colorlinks=true,citecolor=blue,linkcolor=blue]{hyperref}
\def\avg#1{\left\langle#1\right\rangle}

\def\A{\uparrow}
\def\V{\downarrow}

\emergencystretch=\maxdimen
\begin{document}

\title{Spin triplet superconducting pairing in doped MoS$_2$}
\author{Jingyao Wang}
\email{J. Wang and X. Zhang contributed equally to this work.}
\affiliation{Department of Physics, Beijing Normal University,
Beijing 100875, China}
\affiliation{Key Laboratory for Microstructural Material Physics of Hebei Province, School of Science, Yanshan
University, Qinhuangdao 066004, P. R. China}
\author{Xiao Zhang}
\email{J. Wang and X. Zhang contributed equally to this work.}
\affiliation{Department of Physics, Beijing Normal University,
Beijing 100875, China}
\author{Runyu Ma}
\affiliation{Department of Physics, Beijing Normal University,
Beijing 100875, China}
\author{Guang Yang}
\affiliation{Department of Physics, Beijing Normal University,
Beijing 100875, China}
\author{Eduardo V. Castro}
\affiliation{Centro de F\'{\i}sica das Universidades do Minho e Porto, Departamento
de F\'{\i}sica e Astronomia, Faculdade de Ci\^{e}ncias, Universidade do Porto,
4169-007 Porto, Portugal}
\affiliation{Beijing Computational Science Research Center, Beijing 100193, China}\author{Tianxing Ma}
\email{txma@bnu.edu.cn}
\affiliation{Department of Physics, Beijing Normal University,
Beijing 100875, China}
\affiliation{Beijing Computational Science Research Center, Beijing 100193, China}
\date{\today}

\begin{abstract}
Searching for triplet superconductivity has been pursued
intensively in a broad field of material science and quantum information for decades.
Nevertheless, these novel states remain rare. Within a simplified effective three-orbital model, we reveal a spin triplet pairing
in doped MoS$_2$ by employing both the finite temperature determinant quantum Monte Carlo approach and the ground state constrained-phase quantum Monte Carlo method. In a wide filling region of $\avg{n}=0.60-0.80$ around charge neutrality $\avg{n}=2/3$, the $f$-wave pairing dominates over other symmetries. The pairing susceptibility strongly increases as the on-site Coulomb
interaction increases, and it is insensitive to spin-orbit coupling.
\end{abstract}


\maketitle

%
%
{\emph{ Introduction}}:
Currently, physics with electronic states described by the Bloch wave functions
obeying the
Dirac equation has been attracting widespread attention, with graphene and the surface of 3D topological insulators as notable examples \cite{RevModPhys.81.109,RevModPhys.83.1057,vafekVishwanathRev2014}. When combined with superconductivity, the long sought-after Majorana fermions are expected to occur as bound states at vortices in the topological superconducting state \cite{FK08}. The growing interest in realizing unconventional superconductivity, particularly in graphene \cite{PhysRevLett.109.197001,JPCM2014,PhysRevB.92.085121,Nandkishore2012,PhysRevB.90.245114,MaEPL2015}, is justified not only because it pushes us to understand the superconducting mechanisms in solids but also because it may present more opportunities for topological quantum physics \cite{RevModPhys.83.1057}.
In particular, possible triplet superconductivity \cite{PhysRevB.92.085121,Nandkishore2012,MaEPL2015,PhysRevB.90.245114} has been pursued
intensively, as it may be instrumental in the realization of topological quantum computation \cite{Kitaev2001,PhysRevLett.107.217001,Mourik1003,Rokhinson2012,PhysRevLett.109.056803,Anindya2012}.

Recently, MoS$_2$ has been shown to undergo a superconducting transition at high carrier
concentrations when the material is heavily gated to the conducting regime \cite{Ye1193,APL4740268,Lu1353,Saito2016a,Costanzo2016,Piatti2018}.
Interestingly, the Brillouin zone of MoS$_2$ is hexagonal, and around the edges of this zone, the low-energy fermionic excitations behave as massive Dirac particles \cite{xiao2012}.
MoS$_2$ is a monolayer of molybdenum disulfide, which consists of triangularly arranged Mo atoms sandwiched between two layers of triangularly arranged S atoms \cite{PhysRevB.64.235305,nnano2012}.
Ferromagnetic behavior has been reported associated to defects or edges \cite{APL4750237, magnetism2017a, magnetism2017b, magnetism2019}, indicating that the electron-electron interactions in MoS$_2$ are non-negligible.  \cite{Brito2022}. Spontaneous valley polarization has also been detected experimentally \cite{scrace2015magnetoluminescence}, which was attributed to correlated behavior stemming from the transition metal atoms \cite{duanNCommSVP2016, Braz2018}. The combination of electron correlations and massive Dirac physics in MoS$_2$ is expected to lead to novel properties, as for example unconventional superconductivity,
  which could share similarities to that found in doped cuprates or iron-based superconductors \cite{Mazin2010}.

The origin of superconductivity in heavily doped MoS$_2$ has been previously studied, and different superconducting pairing phases have been theoretically suggested \cite{PhysRevB.88.054515,PhysRevB.90.245105,PhysRevLett.113.097001,PhysRevB.93.115406}. For example, depending on the electron-electron interactions and Rashba spin-orbit coupling, MoS$_2$ may show possible topological superconducting phases \cite{PhysRevLett.113.097001}. When both the electron-phonon and electron-electron interactions  are taken into account, some authors report conventional pairing phases \cite{PhysRevB.90.245105}, while others find unconventional superconductivity \cite{PhysRevB.88.054515,PhysRevB.93.115406}.
In this scenario, where electron correlations are non-negligible and different methods point to different pairing mechanisms, it is essential to predict the pairing symmetry using  unbiased, numerically exact tools.
Hartree-Fork type approaches, which have been used widely, are biased if electronic correlations
dominate in the system. Experimentally, recent measurements have excluded a fully gapped superconducting state in MoS$_2$, revealing the presence of a DOS that vanishes linearly with energy \cite{Costanzo2018}. This is a strong indication that a conventional, purely phonon-driven  mechanism is not enough and that electronic correlations do play a role.

There are multiple numerical techniques that have been devoted to the calculations of pairing order parameters. A remarkable theoretical study using  dynamical mean field theory reports
  triplet pairing superconductivity in Sr$_2$RuO$_4$ \cite{Acharya2019}. The possible topological superconducting phases of MoS$_2$ were found using the group theoretical approach \cite{PhysRevLett.113.097001}. By employing the variational Monte Carlo method, the $d$-wave and $p$-wave pairing states on the square lattice have been investigated \cite{FUJITA2007134}. There are also some newly developed methods that embed machine learning techniques to tackle the many-body problem in correlated systems \cite{PhysRevB.104.205120}.
However, due to the rather complex structure of the complete model of MoS$_2$ with eleven orbitals \cite{PhysRevB.88.075409}, numerical tools are difficult to use if one want to treat both the electronic correlations and lattice geometry on the same footing.

In the present work, within a minimum 3-orbital tight-binding model \cite{PhysRevB.88.085433}, we establish spin-triplet superconductivity in doped MoS$_2$ due to electron correlations, using both the finite temperature determinant quantum Monte Carlo (DQMC) \cite{PhysRevD.24.2278,PhysRevB.31.4403,PhysRevB.39.839,PhysRevB.40.506} method and the constrained-phase quantum Monte Carlo (PCPMC) method \cite{MaEPL2015,PhysRevLett.74.3652,PhysRevB.55.7464,1997Stochastic}.
The influence of spin-orbit couplings is also examined.

\begin{figure}[tbp]
\includegraphics[width=8.8cm]{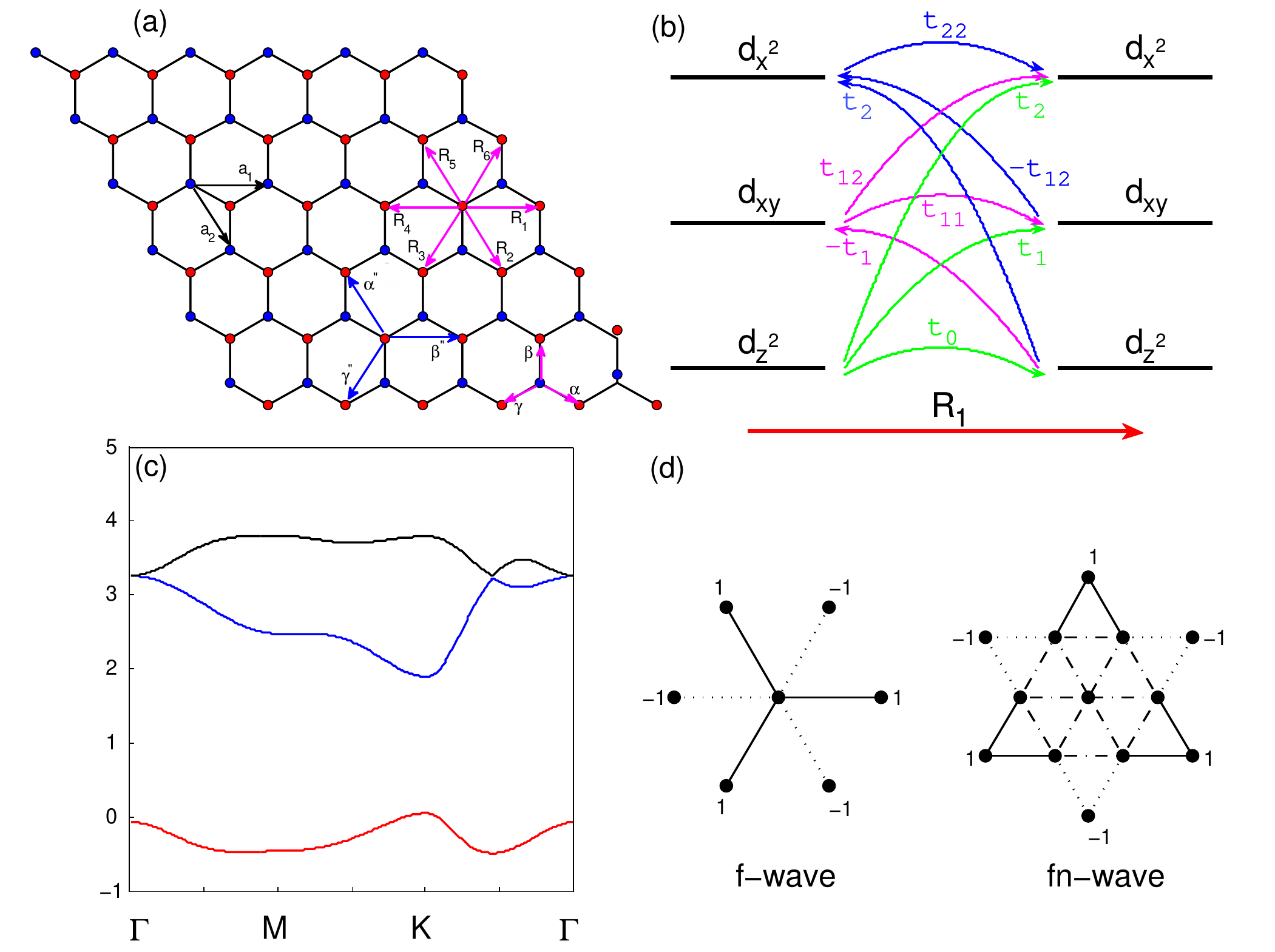}
\caption{(a) MoS$_2$ lattice with indication of unit cell position ${\bf r} = m{\bf a}_1+n{\bf a}_2 $. The vectors connecting nearest-neighbors are $\alpha,\beta, \gamma$, while those connecting
  next-nearest-neighbors
  are $\alpha', \beta', \gamma'$ (or ${\bf R}_1, {\bf R}_2, ..., {\bf R}_6$).
  (b) Hopping integrals between two transition metal atoms connected by the lattice vector ${\bf R}_1$. (c) Electronic structure of MoS$_2$ for the 3-orbital tight binding model with $t_0=-1.0, t_1=2.2, t_2=2.3, t_{11}=1.3, t_{12}=1.6, t_{22}=0.16, \epsilon_1=4.6$ and $\epsilon_2=10.5$ in units of $|t_0|$. 
  (d) Site-dependent form factors for $f$-wave and $fn$-wave pairing in the triangular lattice.}
\label{Fig:Sketch}
\end{figure}

{\it Model and methods:} To make use of highly
controllable and unbiased numerical methods to study the physical properties caused by electronic correlations in MoS$_{2}$,
a simplified Hamiltonian is required. Next, we briefly introduce the effective 3-orbital tight binding model for MoS$_{2}$\cite{PhysRevB.88.085433}, which contains two key ingredients for this material: $d-$orbitals coming from transition metal atoms and the concomitant massive Dirac physics at low-energies.

Essentially, this material forms an hexagonal lattice with three atoms per unit cell: one transition metal and two chalcogens,
and its lattice structure is shown in Fig.~\ref{Fig:Sketch}(a). The honeycomb structure occurs only in the so-called $2H$ phase, which is the phase of MoS$_2$.
The lattice is composed of three layers:
the top and bottom  layer made of chalcogen atoms, and a the middle layer, half way between these two, made of molybdenum atoms  shown as red circles in Fig.~\ref{Fig:Sketch}(a).
The two chalcogens, shown as blue circles in Fig.~\ref{Fig:Sketch}(a), are vertically aligned and sit on top of each other.

The simplest tight binding model for this material \cite{PhysRevB.88.085433}
considers only
three $d-$orbitals sitting at the sites of  the triangular lattice.
As shown in Fig.~\ref{Fig:Sketch}(b), we can define six hopping parameters, which, following the notation of Ref.~\cite{PhysRevB.88.085433}, can be denoted by $t_0, t_1, t_2, t_{11}, t_{22}, t_{12}$. In the figure, we indicate both forward and backward hoppings (see the arrows). Note that for hopping integrals involving the orbitals $d_{xy}$, there is a negative sign difference between forward and backward hoppings due to the symmetry of the $d_{xy}$ orbital.

The model we consider in this work is the 3-orbital tight binding model on a triangular lattice with Hubbard like on-site interaction.  The Hamiltonian may be written as $H =H_0+H_U$, where for the kinetic part $H_0$, the complication comes from the fact that we have three orbitals per lattice site and six different direction-dependent hoppings between the
three orbitals. In the basis $d_{z^2},d_{xy},d_{x^2 - y^2}$, the matrix elements of $H_0$ can be written in direct space as $3 \times 3$ matrices,
\begin{equation}
\mathcal{H}_{\bf r,r}=\left[
\begin{array}{ccc}
\epsilon_1 & 0 & 0 \\
0 & \epsilon_2 & 0 \\
0 & 0 & \epsilon_2 \\
\end{array}
\right],\label{Hrr}
\end{equation}

\begin{equation}
\mathcal{H}_{\bf r,r\pm{a_1}}=\left[
\begin{array}{ccc}
t_0 & \pm{t_1} & t_2 \\
\mp{t_1} & t_{11} & \pm{t_{12}} \\
t_2 & \mp{t_{12}} & t_{22} \\
\end{array}
\right],
\end{equation}

\begin{widetext}
\begin{equation}
\mathcal{H}_{\bf r,r\pm{a_2}}=\left[
\begin{array}{ccc}
t_0 & \pm{\frac{1}{2}}{t_1}-{\frac{\sqrt3}{2}}{t_2} & \mp{\frac{\sqrt3}{2}}{t_1}-{\frac{1}{2}}{t_2} \\
\mp{\frac{1}{2}}{t_1}-{\frac{\sqrt3}{2}}{t_2} & {\frac{1}{4}}t_{11}+{\frac{3}{4}}t_{22} & {\frac{\sqrt3}{4}}[t_{22}-t_{11}]\mp{t_{12}} \\
\pm{\frac{\sqrt3}{2}}{t_1}-{\frac{1}{2}}{t_2} & {\frac{\sqrt3}{4}}[t_{22}-t_{11}]\pm{t_{12}} & {\frac{3}{4}}t_{11}+{\frac{1}{4}}t_{22} \\
\end{array}
\right],
\end{equation}

\begin{equation}
\mathcal{H}_{\bf r,r\pm{(a_2-a_1)}}=\left[
\begin{array}{ccc}
t_0 & \pm{\frac{1}{2}}{t_1}+{\frac{\sqrt3}{2}}{t_2} & \mp{\frac{\sqrt3}{2}}{t_1}-{\frac{1}{2}}{t_2} \\
\mp{\frac{1}{2}}{t_1}+{\frac{\sqrt3}{2}}{t_2} & {\frac{1}{4}}t_{11}+{\frac{3}{4}}t_{22} & -{\frac{\sqrt3}{4}}[t_{22}-t_{11}]\mp{t_{12}} \\
\pm{\frac{\sqrt3}{2}}{t_1}-{\frac{1}{2}}{t_2} & -{\frac{\sqrt3}{4}}[t_{22}-t_{11}]\pm{t_{12}} & {\frac{3}{4}}t_{11}+{\frac{1}{4}}t_{22} \\
\end{array}
\right].\label{HrrpF}
\end{equation}
\end{widetext}
The band structure for  $H_0$ along path $\Gamma-M-K-\Gamma$ in the hexagonal Brillouin zone is shown in Fig.~\ref{Fig:Sketch}(c). Charge neutrality occurs when the lowest band is totally filled, with a number of electrons per unit cell and orbit of $\avg{n}$=2/3. The finite band gap to the next band makes these materials semiconducting.
Around the $K$-point, where the low energy massive Dirac spectrum appears, the result is in good agreement with the $11$-orbital model\cite{PhysRevB.88.075409, Silva-Guillen2016}. Working in units of $|t_0|$,
the parameters used here are $t_0=-1.0, t_1=2.2, t_2=2.3, t_{11}=1.3, t_{12}=1.6, t_{22}=0.16, \epsilon_1=4.6$ and $\epsilon_2=10.5$. These parameters are representative also for other members of the transition metal dichalcogenides family~\cite{PhysRevB.88.085433}, like WSe$_2$ (as a reference, $t_0 =-0.184\,$eV for MoS$_2$ and $t_0 =-0.207\,$eV for WSe$_2$). In the following we use $t\equiv|t_0|$ as the energy unit.

A distinctive aspect of MoS$_{2}$ is that, unlike graphene, the spin-orbit coupling in these material cannot be neglected\cite{RevModPhys.81.109}.
Actually, the spin-valley coupling in transition metal dichalcogenides, which opens the door to the possible control of spin and valley degrees of freedom \cite{xiao2012, nphys2014}, is rooted in the non negligible spin-orbit coupling of these materials.
The origin of the spin-orbit coupling in MoS$_{2}$ is
due to the heavy transition metal atoms, and is well described by
the intrinsic contribution
$H_{SO}=\lambda {\bf L}\cdot{\bf S}$ \cite{xiao2012, PhysRevB.88.085433}.
It gives rise to a characteristic spin-splitting of the valence band maximum, which has been measured in excellent agreement with theory \cite{socExp2015}. The value of the spin-orbit coupling parameter $\lambda$ depends on the specific transition metal atom, but is found to be in the range $\lambda \approx t$ (according to Ref.~\cite{PhysRevB.88.085433}, for MoS$_2$ it is  $\lambda \simeq 0.40t$ and for WSe$_2$ it is $\lambda \simeq 1.1t$).
Within first-order perturbation theory, we can readily
identify the matrix elements of the spin-orbit Hamiltonian in
our tight-binding approach,
\begin{equation}
H^{(spin)}_{\bf{r,r}}=\left[
\begin{array}{cc}
\mathcal{H}_{\bf{r,r}}+\frac{\lambda}{2}L_z &   0  \\
0 &  \mathcal{H}_{\bf{r,r}}-\frac{\lambda}{2}L_z  \\
\end{array}
\right]
\end{equation}
where $\mathcal{H}_{\bf r,r}$ is given by Eq.~\eqref{Hrr} and the matrix ${L_z}$ is the z component of the orbital angular momentum,
\begin{equation}
L_{z}=\left[
\begin{array}{ccc}
0 &   0 &  0 \\
0 &   0 & 2i \\
0 & -2i &  0 \\
\end{array}
\right].
\end{equation}

The simplified effective 3-orbital tight binding model
 may be seen as a three-layer triangular lattice; one layer for each of the three $d-$orbitals in the model.
For these
types of structures, one can design a lattice with $3 \times 3 \times L^2$ sites,
which could be reachable by using the finite temperature DQMC and PCPMC methods for $L\leq5$.
Thus, the effective simplified model provides a new opportunity to make use of highly
controllable and unbiased numerical methods to study the possible
electronic correlation-driven phases
in transition metal dichalcogenide materials, in particular superconductivity.

From the simplified effective 3-orbital tight binding model, the Hamiltonian can be written as
\begin{eqnarray}
H&=&\sum_{\left  \langle {\bf {r,r^\prime}}\right\rangle, \sigma}c^{\dagger}_{{\bf r} \sigma}\mathcal{H}_{\bf r,r^\prime} c_{{\bf r^\prime} \sigma}
  +\sum_{{\bf r} , \sigma}c^{\dagger}_{{\bf r} \sigma}\left (\mathcal{H}_{{\bf r,r}} + \frac{\lambda}{2}\sigma L_z\right )c_{{\bf r} \sigma}  \nonumber \\
&+&U\sum_{{\bf r},\gamma }n_{{\bf r} \gamma \uparrow}n_{{\bf r} \gamma \downarrow} - \mu\sum_{{\bf r}, \gamma, \sigma}n_{{\bf r} \gamma \sigma }\label{H}
\end{eqnarray}
where
$c^\dag_{{\bf r} \sigma} = (c^\dag_{{\bf r}, d_{z^2}, \sigma}, c^\dag_{{\bf r}, d_{xy}, \sigma}, c^\dag_{{\bf r}, d_{x^2-y^2}, \sigma})$ is a row vector in orbital space and  $c^\dag_{{\bf r}, \gamma, \sigma}$
is the electron creation operator at lattice site $\bf r$ of the effective layer (orbital) $\gamma$
and with spin polarization $\sigma=\A,\V$. The third  term is $H_U$, where
$U$ labels the on-site repulsive interaction, and in the last term $\mu$ is the chemical potential.
  The explicit expressions for the $\mathcal{H}_{\bf r,r^\prime}$ matrices, with $\bf r,r^\prime$ nearest neighbors, are given in Eqs.~\eqref{Hrr}-\eqref{HrrpF}.

The numerical method used is the DQMC approach and PCPMC methods.
The DQMC approach has been widely used for decades \cite{PhysRevD.24.2278,PhysRevB.31.4403,PhysRevB.39.839,PhysRevB.40.506,PhysRevB.84.121410,PhysRevLett.110.107002,PhysRevB.90.245114,PhysRevB.101.155413,PhysRevB.104.035104}.
The PCPMC method is a generalization of the constrained-path method (CPMC)\cite{PhysRevLett.74.3652,PhysRevB.55.7464,1997Stochastic} and is an analog of the fixed-phase generalization of the fixed-node diffusion Monte Carlo method \cite{PhysRevLett.71.2777}. The PCPMC method has yielded very accurate results for the ground state energy and other ground state observables for various strongly correlated lattice models \cite{PhysRevLett.104.116402,PhysRevB.78.165101,PhysRevA.82.061603,PhysRevB.105.155154} and for atoms, molecules, and nuclei \cite{SCHMIDT199999}.


The constrained-path method approximately handles the sign problem, which is caused by a broken symmetry in the space of Slater determinants, by eliminating any random walker as soon as $\langle\phi_i|\psi_T\rangle < 0$. The presence of the spin-orbit interaction in the Hamiltonian means that the ground state cannot be real. To ensure that samples come from a real, non-negative distribution, the constrained-phase approximation generalizes the constrained-path condition: with a phase $\theta$ defined by
\[
e^{i\theta} \equiv \langle\phi|\psi_T \rangle / |\langle\phi|\psi_T\rangle|,
\]
two simple forms of the constrained-phase method follow \cite{PhysRevLett.74.3652} from replacing the walker either by
$
|\phi\rangle \leftarrow \cos(\theta) e^{-i\theta}|\phi_i\rangle
$,
and eliminating the walker if $\Re{\langle\phi_i|\psi_T\rangle} < 0$, or by
$
|\phi\rangle \leftarrow e^{-i\theta}|\phi_i\rangle
$,
which makes $\langle\phi_i|\psi_T\rangle > 0$.
Here, we used the first constraint.
In the PCPMC method, extensive benchmark calculations showed that the systematic
error induced by the constraint is within a few percent and the
ground-state observables are insensitive to the choice of trial
wave function. In our PCPMC simulations, we employ closed-shell
electron fillings and use the corresponding free-electron
$U$=0 wave function as the trial wave function.
To further justify the accuracy of our PCPMC method, we provide a comparison between PCPMC and Exact Diagonalization methods in the Appendix~\ref{appendix}.
For more details, we refer to Refs.~\cite{MaEPL2015,SCHMIDT199999,PhysRevC.68.024308}.

To investigate the superconducting properties, we compute the pairing susceptibility \cite{PhysRevB.39.839,PhysRevB.40.506,PhysRevB.84.121410,PhysRevLett.110.107002,PhysRevB.90.245114,PhysRevB.101.155413,PhysRevB.104.035104},
\begin{equation}
P_{\alpha}=\frac{1}{N_s}\sum_{\bf r,r^\prime,\gamma}\int_{0}^{\beta }d\tau \langle \Delta
_{\alpha }^{\dagger }({\bf r,\gamma},\tau)\Delta _{\alpha }^{\phantom{\dagger}%
}({\bf r^\prime,\gamma},0)\rangle,\label{sus}
\end{equation}
 where  $\Delta_{\alpha }^{\dagger }(\bf r,\gamma)$ is the corresponding order parameter, written as
\begin{eqnarray}
\Delta_{\alpha }^{\dagger }({\bf r},\gamma)\ =\sum_{m}f_{\alpha}^{\dagger}
(\delta_{m})(c_{{\bf r \gamma}\uparrow }c_{{{\bf r+\delta}_{m} \gamma}\downarrow } {\color{blue} \pm }
c_{{\bf r \gamma}\downarrow}c_{{{\bf r+\delta}_{m} \gamma}\uparrow })^{\dagger}. \label{Delta}
\end{eqnarray}
Here, $\alpha$ stands for the pairing symmetry, $f_{\alpha}({\bf \delta}_{m})$ is the form factor of the pairing
function, and the vectors ${\bf \delta}_{m}$ connect
nearest neighbor  $(m=1,2,...,6)$ or next-to-nearest neighbor $(m=1,2,...,12)$ sites. We specify below the pairing symmetries we consider in this work.

The pairing correlation function we compute is given by
\begin{eqnarray}
C_{\alpha}({\bf r}={\bf r}_{\bf i}-{\bf r}_{\bf j})=\sum_{\gamma}\langle\Delta_{\alpha}^{\dagger}({\bf r_i,\gamma})\Delta_{\alpha}^{\phantom{\dagger}}({\bf r_j,\gamma})\rangle.
\end{eqnarray}
To extract the intrinsic pairing correlation in the finite system, we also examined the vertex contributions to the correlations defined by
\begin{eqnarray}
V_{\alpha}(\bf{r})=\emph{C}_{\alpha}(\bf{r})-\overline{\emph{C}_{\alpha}}(\bf{r}), \label{vertex}
\end{eqnarray}
where $\overline{\emph{C}_{\alpha}}(\bf{r})$ represents the uncorrelated single-particle contribution. Each term in $C_{\alpha}(\bf{r})$, such as $\langle a^{\dagger}_{\uparrow}a^{\phantom{\dagger}}_{\uparrow}a^{\dagger}_{\downarrow}a^{\phantom{\dagger}}_{\downarrow} \rangle$, has a corresponding term $\langle a^{\dagger}_{\uparrow}a^{\phantom{\dagger}}_{\uparrow} \rangle\langle a^{\dagger}_{\downarrow}a^{\phantom{\dagger}}_{\downarrow} \rangle$.

{\it Results and discussion:}
The basic geometry in the structure of the simplified model is a three-layer triangular
lattice.
For the superconductivity in MoS$_2$, the possible dominant pairing symmetry in one plane of the
 triangular
lattice may play a key role, as in doped cuprates for multilayer superconductors\cite{RevModPhys.78.17}.
In a triangular
lattice, we may consider
seven
 types of pairing forms, with form factors
  $f_{s}(l)$, $f_{d}(l)$, and $f_{d_n}(l)$, for the singlet pairing, and $f_{p}(l)$, $f_{p_n}(l)$, $f_{f}(l)$, and $ f_{f_n}(l)$,  for the triplet pairing
  (see Refs.~\cite{Ogata-PhysRevB.72.134513,JPCMWu2013,Cheng2014}). The subscript $n$ in $p_n$-, $d_n$-, and $f_n$-wave symmetry, refers to the next-to-nearest neighbor version of Eq.~\eqref{Delta}.
Among those
we study,
 the $f$- and
$f_n$-wave pairing factors are
shown in Fig.\ref{Fig:Sketch}(d), which are both triplet pairing forms. 

\begin{figure}[tbp]
\includegraphics[width=8.8cm]{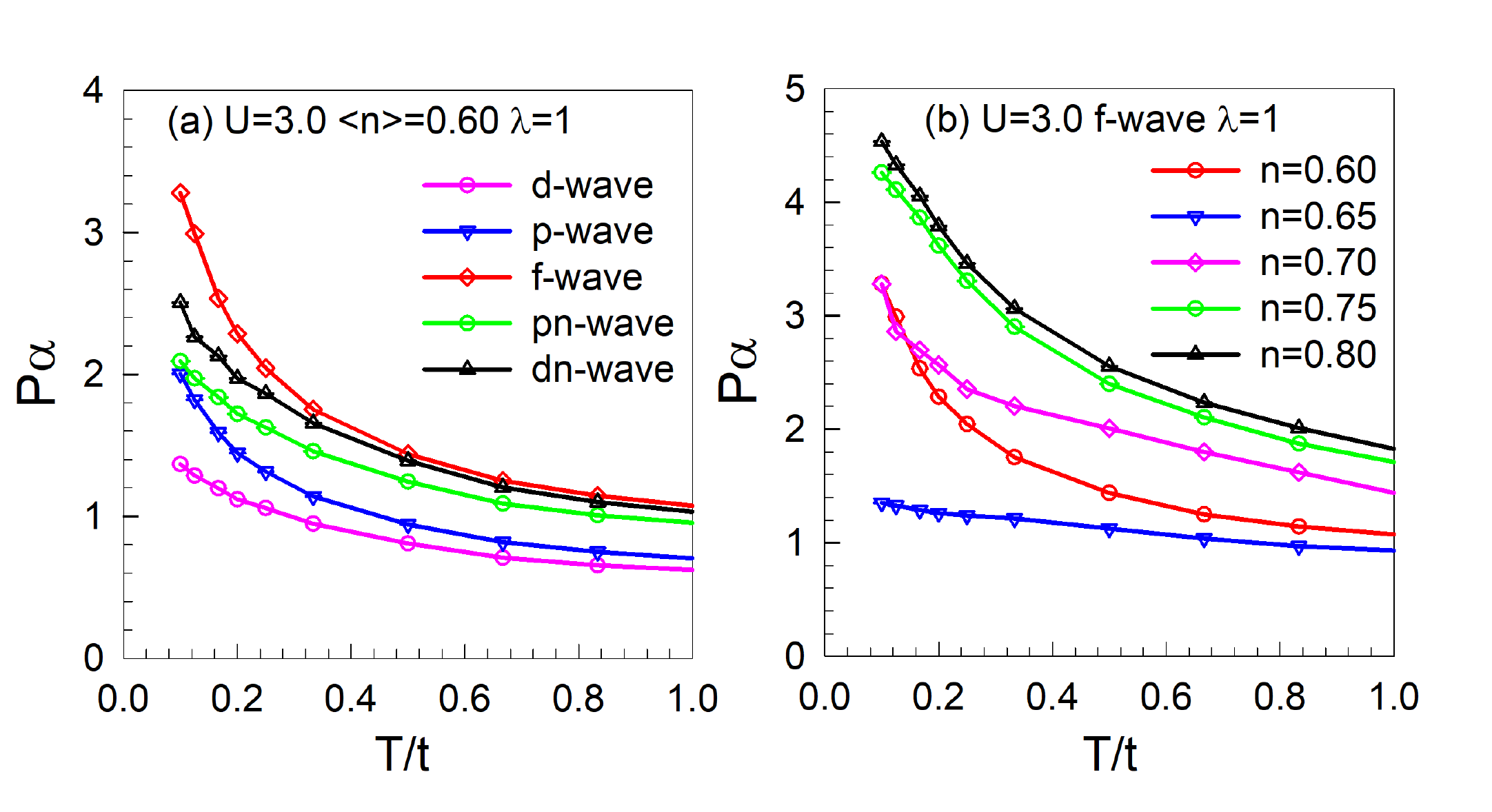}
\caption{(a) Pairing susceptibility $P_{\alpha}$ for different pairing symmetries versus temperature $T$ at $U=3.0t$, $\avg{n}$=0.60 and ${\lambda}=1.0t$. (b) $f$-wave pairing susceptibility versus  temperature $T$ at $U=3.0t$ and ${\lambda}=1.0t$ for different electronic fillings around charge neutrality.}
\label{Fig:Pn0608}
\end{figure}

In Fig.~\ref{Fig:Pn0608}(a), we present the pairing susceptibility with different symmetries as a function of temperature at the electron filling $\avg{n}$=0.6 for $U=3.0t$ and spin-orbit coupling parameter $\lambda=1.0t$. It is interesting to see that for the electronic filling we investigated, triplet pairing with $f$-wave symmetry dominates.
Moreover, as the temperature decreases, the pairing susceptibility increases, which means that the pairing susceptibility
may diverge at some low temperature, resulting in the existence of superconductivity.
Fig.~\ref{Fig:Pn0608}(b) shows the pairing susceptibility with $f$-wave symmetry
 for different electronic fillings around charge neutrality.  Recall that charge neutrality occurs at $\avg{n}$=2/3, when the lowest band
shown in Fig.~\ref{Fig:Sketch}(c)  is completely filled.
 In Fig.~\ref{Fig:Pn0608}(b), for $\avg{n}=0.65$ the pairing susceptibility hardly increases as $T$ is lowered. In this case, the system is very close to the band insulating state. As either the hole or electron density is increased, the pairing susceptibility also increases at lower temperatures. This agrees with experiments, where high carrier densities are required for superconductivity to be observed.

\begin{figure}[tbp]
\includegraphics[width=8.5cm]{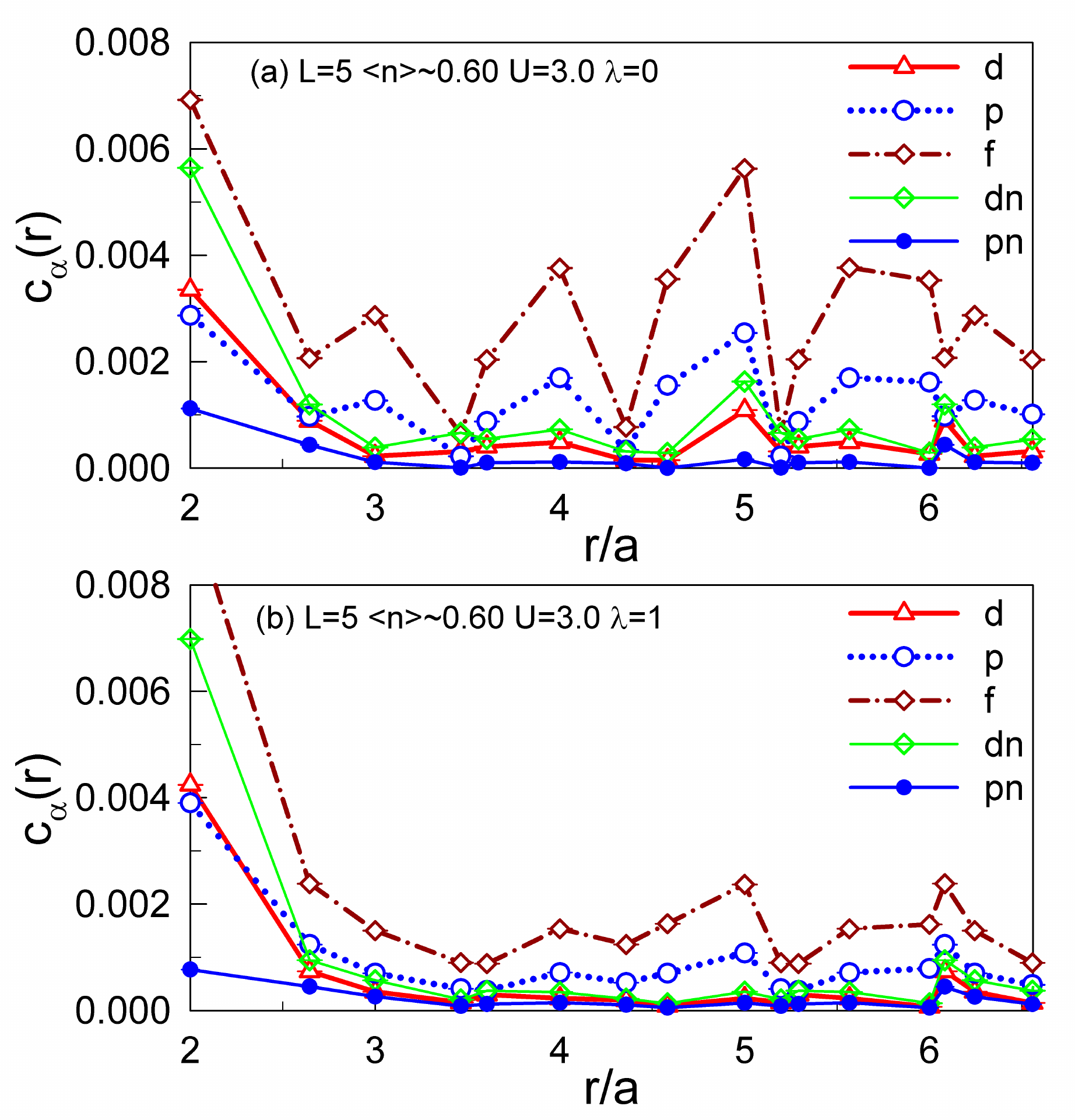}
\caption{The pairing correlation function for different pairing symmetries versus the distance $r$ at $U =3.0t$ and $\avg{n}\approx 0.60$ with (a) ${\lambda}$=0 and (b) ${\lambda}=1.0t$. Here $a$ refers the lattice spacing.
}
\label{Fig:soc}
\end{figure}

To further identify which pairing symmetry is dominant
 through numerical calculations
 in finite size
 systems, we investigate the long-range part of the ground state pairing correlation
function \cite{PhysRevLett.70.682,PhysRevLett.78.4486,PhysRevB.84.121410}
 using the PCPMC method.
In Fig.~\ref{Fig:soc}, we compare the long-range part of the pairing correlation  function for different pairing symmetries. In Fig.~\ref{Fig:soc}(a) we show the results obtained in the absence of spin-orbit coupling, $\lambda$=0, and in Fig.~\ref{Fig:soc}(b) for a finite spin-orbit coupling of $\lambda=t$.
The simulations were performed on the $L$=5 lattice at $U=3.0t$ and $\avg{n}$$\approx$0.60, with a closed shell filling of $N_{up}=N_{dn}=67$.
 As can be seen in Fig.~\ref{Fig:soc}(a),
when there is no spin-orbit coupling, the $f$-wave pairing symmetry plays a dominant role. When the spin-orbit coupling increases to $\lambda = 1.0t$ in Fig.~\ref{Fig:soc}(b), the $f$-wave pairing symmetry still dominates
for all long-range distances between electron pairs.
When the spin-orbit coupling increases from $\lambda = 0$ to $\lambda=1.0t$, the pairing correlation function for
$d$-, $p$-, $f$-, and $d_n$-wave symmetry decreases, while that for the $p_n$-wave always stays close to zero.
 We conclude that while an increased spin-orbit coupling
 may inhibit the pairing correlations, it
does not affect the main electron pairing forms.
\begin{figure}[tbp]
\includegraphics[width=8.5cm]{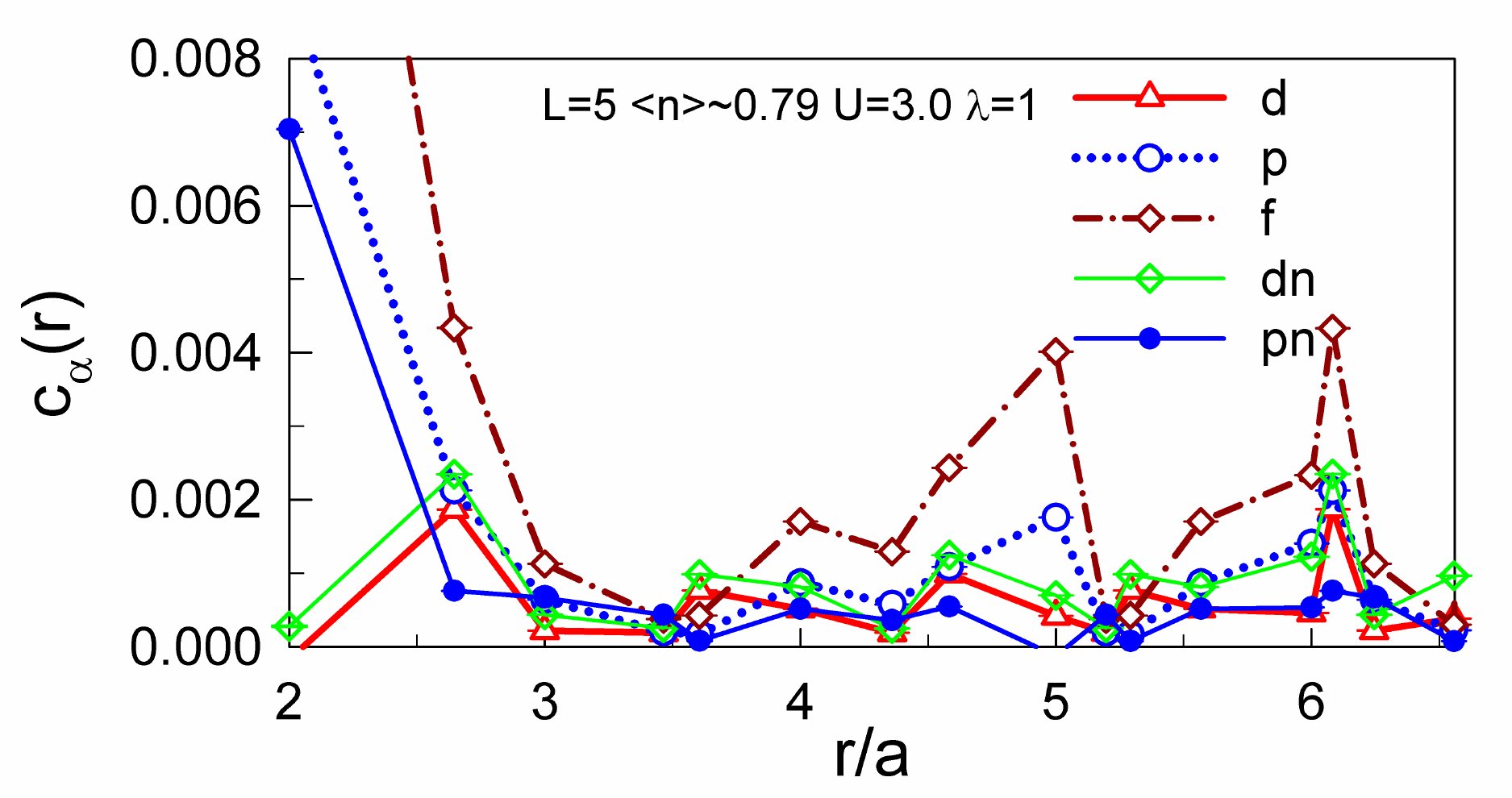}
\caption{The pairing correlation function for different pairing symmetries versus the distance $r$ at $U=3.0t$ and $\avg{n}\approx 0.79$ with ${\lambda}=1.0t$. }
\label{Fig:density}
\end{figure}

To study the influence of the electron density on the pairing correlations function $C_{\alpha}$, we increased the electron density to $\avg{n}\approx 0.79$, which corresponds to a closed shell filling of $N_{up}=N_{dn}=89$. The results are shown in Fig.~\ref{Fig:density}. Appart from a different electron density, we used the same parameters as those in Fig.~\ref{Fig:soc}(b). It can be seen that by increasing the electron density to $\avg{n}\approx0.79$ increases the $C_{\alpha}$ of all pairing symmetries. Nevertheless, the $f$-wave symmetry still dominates over other pairing symmetries, which points to an $f$-wave superconducting state. This also reinforces our findings in Fig.\ref{Fig:soc}.


\begin{figure}[tbp]
\includegraphics[width=8.5cm]{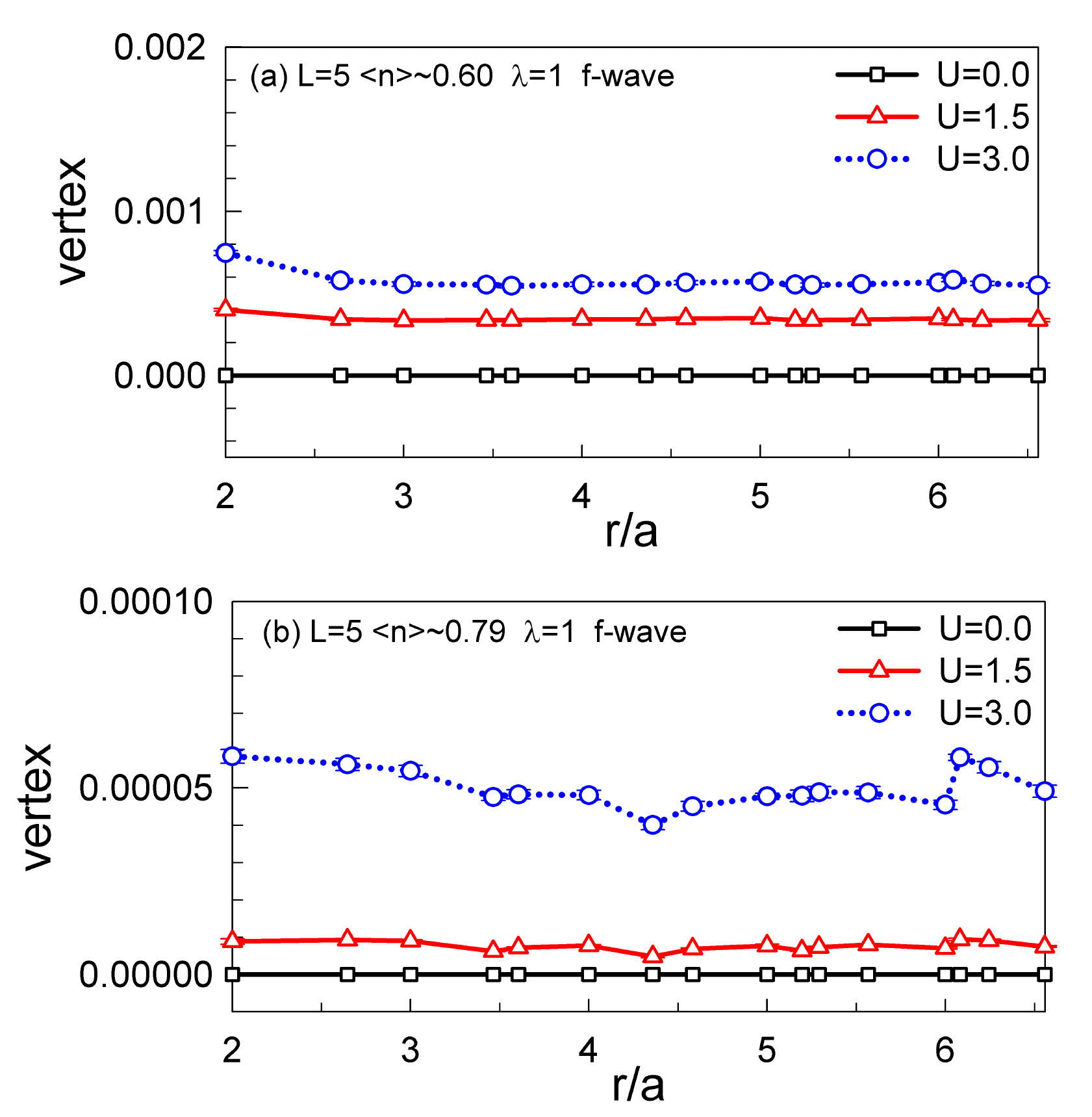}   
\caption{The vertex contribution to the pairing
correlation function for different interactions $U$ versus the distance $r$ at $\avg{n}\approx 0.60$ (a)  and $\avg{n} \approx 0.79$ (b).  In both panels we set
    ${\lambda}=1.0t$ and show the results for $U$=0.0, $U=1.5t$, and $U=3.0t$. }
\label{Fig:vertex}
\end{figure}

In Fig.~\ref{Fig:vertex} we show the vertex contribution
to the pairing correlation function, given by Eq.~\eqref{vertex},
for the $f$-wave pairing symmetry at zero temperature. In panel~\ref{Fig:vertex}(a) the electron density was set to
$\avg{n}=0.60$ and in panel~\ref{Fig:vertex}(b) to $\avg{n}=0.79$, both with a spin-orbit coupling $\lambda = 1.0t$.
 The results are shown at three different values of the interaction: $U=0.0$, $1.5t$, and $3.0t$. It is clear
that the vertex contribution becomes larger when the interaction increases
Comparing Fig.~\ref{Fig:vertex}(a) and Fig.~\ref{Fig:vertex}(b), we find
that an increased electron density makes the vertex contribution smaller.

\begin{figure}[tbp]
\includegraphics[width=8.5cm]{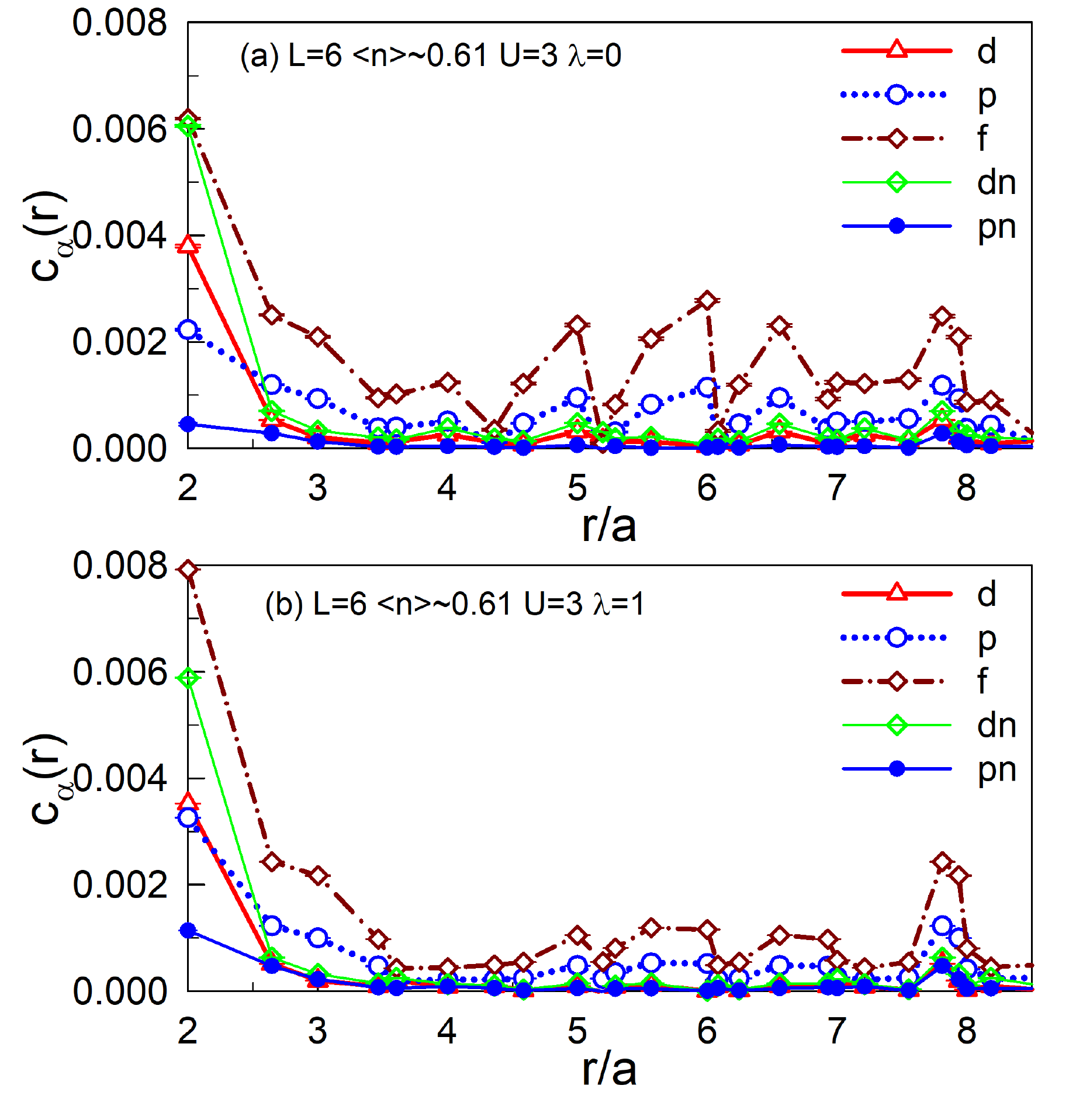}   
\caption{The pairing correlation function for different pairing symmetries versus the distance $r$ on $L=6$ lattice at $U = 3.0t$ and $\avg{n}\approx 0.61$, with a closed-shell filling of $N_{up} $$=$$N_{dn}$$=$99. The spin-orbit coupling is (a) ${\lambda}$=0 and (b) ${\lambda}=1.0t$.}
\label{Fig:size}
\end{figure}

In this work, we focus on the close-shell case,
which leads us to naturally choose the corresponding free-electron wave function as the trial wave function.
The selection of close-shell filling can generally reflect the physical properties of the doped system.
To further verify our results, Fig. \ref{Fig:size} shows simulations performed on a larger lattice of $L$=6 at $U$=3.0$t$ and $\avg{n}$$\approx$0.61, with a close-shell filling of $N_{up} $$=$$N_{dn}$$=$99. By comparing Fig. \ref{Fig:soc} and Fig.\ref{Fig:size}, it can be seen that in the $L$=6 lattice, it remains the same as in the $L$=5 lattice that the $f$-wave pairing symmetry still dominates among all the electron pairing symmetries.
These data provides evidence that the finite size effect caused by close-shell filling does not affect the qualitative results we are concerned with, where we need not to make a filling dependent systematic finite-size study.

{\emph{{Conclusions:}} By using \emph{finite temperature determinant quantum Monte Carlo}
and the \emph{constrained-phase quantum Monte Carlo} methods, we revealed spin-triplet superconducting pairing in doped MoS$_2$ within a simplified effective 3-orbital model. Our intensive unbiased numerical results show that in a wide electron filling region, superconducting pairing with $f$-wave symmetry dominates, regardless of the spin-orbit couplings. The enhanced pairing susceptibility with increasing interaction indicates that, in a scenario where electronic correlations dominate over the electron-phonon coupling, unconventional triplet superconductivity is to be expected in doped MoS$_2$.

{\emph{ Acknowledgments:}} 
T. Ma thanks CAEP for partial financial support. We thank Ming-Qiao Ai for helpful discussion. This work is supported in part by NSCFs (Grant Nos. 11974049). EVC acknowledges funding from Funda ção para a Ci ência e Tecnologia (FCT-Portugal) through Grant No. UIDB/04650/2020. We also acknowledge the computational support from the Beijing Computational Science Research Center (CSRC) and the support from the HSCC of Beijing Normal University.

\appendix

\setcounter{equation}{0}
\setcounter{figure}{0}
\renewcommand{\theequation}{A\arabic{equation}}
\renewcommand{\thefigure}{A\arabic{figure}}
\renewcommand{\thesubsection}{A\arabic{subsection}}

\section{Benchmarking the PCPMC method}
\label{appendix}
To justify the accuracy of the PCPMC method, here we present  results that
confirm the reliability of this computational method.

In Tab.~\ref{Table} we show a comparison of the PCPMC method with Exact Diagonalization (ED)
 results on  the square lattice with different electron dopings and spin orbit interaction strengths.
 The PCPMC method agrees well with the ED results for the energies, double occupancies, and spin-spin correlations.
We also make a comparison of the PCPMC method with projective auxiliary-field quantum Monte Carlo (PQMC) algorithm on the Kane-Mele-Hubbard model, and at half-filling, where there is no sign problem in the corresponding PQMC simulations.
The PQMC algorithm used for the simulations
constitutes an unbiased, controlled and numerically exact method~\cite{SUGIYAMA19861}.

\begin{widetext}
    \begin{center}
    \begin{table}[h]
    \caption{Comparisons of the energies (total $E_T$), double occupancies and correlation
    functions ($S$ and $S_d$ for the spin and charge density structure factors ) of the two-dimensional $U$=4.0 Hubbard
    model on different lattices.
    Different rows correspond to ED and PCPMC results, or PQMC and PCPMC results. The multi-column heading of $4\times 4$ lattice with $ 5\uparrow\,5\downarrow$ refers to the Hubbard model on a square lattice, that of $4\times 4$ lattice with $ 4\uparrow\,4\downarrow$ to the Hubbard model on a square lattice in a magnetic field with 1/4 of flux quanta per plaquette, and that of $2\times 3^2$ KMH model with $ 9\uparrow\,9\downarrow$ to the Kane-Mele Hubbard model with $\lambda=0.1$.}\label{Table}
    \begin{tabular}{|c|c|c|c|c||c|c|c||c|c|c|c|c|c}
    \hline
        && \multicolumn{3}{c||}{ $4\times 4$ lattice with $ 5\uparrow\,5\downarrow$ }
        & \multicolumn{3}{c||}{ $4\times 4$ lattice with $ 4\uparrow\,4\downarrow$ with a flux }
        & \multicolumn{4}{c|}{$2\times 3^2$ KMH model with $ 9\uparrow\,9\downarrow$}\\ \cline{3-11}
     \hline
        &           & $E_T$     & S$(\pi,\pi)$ & S$_{d}(\pi,\pi)$  & $E_T$         & DOC               & S$(\pi,\pi)$  &       & $E_T$         & DOC           & S$_z$ (1,2)   \\
    \hline																						
    U=1.0&	ED	    &-22.568   	&	0.6637   	&	-0.07641	   & -21.712  	&	0.052227	&	0.5196	    &PQMC	&	-24.58(1)	&	0.2260(1)	&	-0.1389(3)	\\
         &	PCPMC	&-22.570(2)	&	0.6636(3)	&	-0.07645(5)	   & -21.713(2)	&	0.052246(7)	&	0.5195(1)	&	PCPMC	&	-24.57(1)	&	0.2263(1)	&	-0.1397(7)	\\
    \hline																						
    U=2.0&	ED	    &-21.377	&	0.6943  	&	-0.08205       & -20.95	    &	0.043493	&	0.5345		&PQMC	&	-20.74(2)	&	0.2009(5)	&	-0.1572(1)	\\
         &	PCPMC	&-21.380(3)	&	0.6943(4)	&	-0.08219(8)	   & -20.950(3)	&	0.043535(1)	&	0.5342(2)	&	PCPMC	&	-20.70(2)	&	0.2021(2)	&	-0.1573(1)	\\
    \hline																						
    U=3.0&	ED	    &-20.392	&	0.717    	&	-0.08702	   & -20.312    &	0.036196	&	0.5454		&PQMC	&	-17.36(7)	&	0.1744(1)	&	-0.1776(3)	\\
         &	PCPMC	&-20.395(5)	&	0.7171(7)	&	-0.08711(1)	   & -20.315(4)	&	0.036251(1)	&	0.5452(3)	&	PCPMC	&	-17.26(3)	&	0.1760(2)	&	-0.1801(2)	\\
    \hline																						
    U=4.0&	ED	    &-19.581	&	0.7327	    &	-0.09115	   & -19.783  	&	0.030185	&	0.5528		&PQMC	&	-14.48(8)	&	0.1434(1)	&	-0.2066(4)	\\
         &	PCPMC	&-19.581(2)	&	0.7336(7)	&	-0.09130(2)	   &-19.775(7)	&	0.030199(2)	&	0.5528(5)	&	PCPMC	&	-14.30(5)	&	0.1484(4)	&	-0.1959(4)	\\
    \hline																						
    \hline

    \hline
    \end{tabular}
    \end{table}
    \end{center}
\end{widetext}
For the three cases shown in Table~I, the key point is that the PCPMC method agrees well with both ED and PQMC results, thus allowing for accurate simulations.

Figure.~\ref{Fig:ED} shows the 4$\times$4 sites Hubbard model in a magnetic field of 1/4 flux quanta per plaquette with four spins
up (and four down) electrons. The potential energy is shown in panel~\ref{Fig:ED}(a)  and  the double occupancy in panel~\ref{Fig:ED}(b), both  as a function of
$U$. The red line with circles is for ED and  blue dots for  PCPMC. The error bar is within the symbols.
One can see that the free-electron $U=0$ wave function tends to be a good choice for $|\Psi_{T}\rangle$, even when  $U$ is increased up to the band-width of square lattice.

\begin{figure}[tbp]
\includegraphics[width=8.5cm]{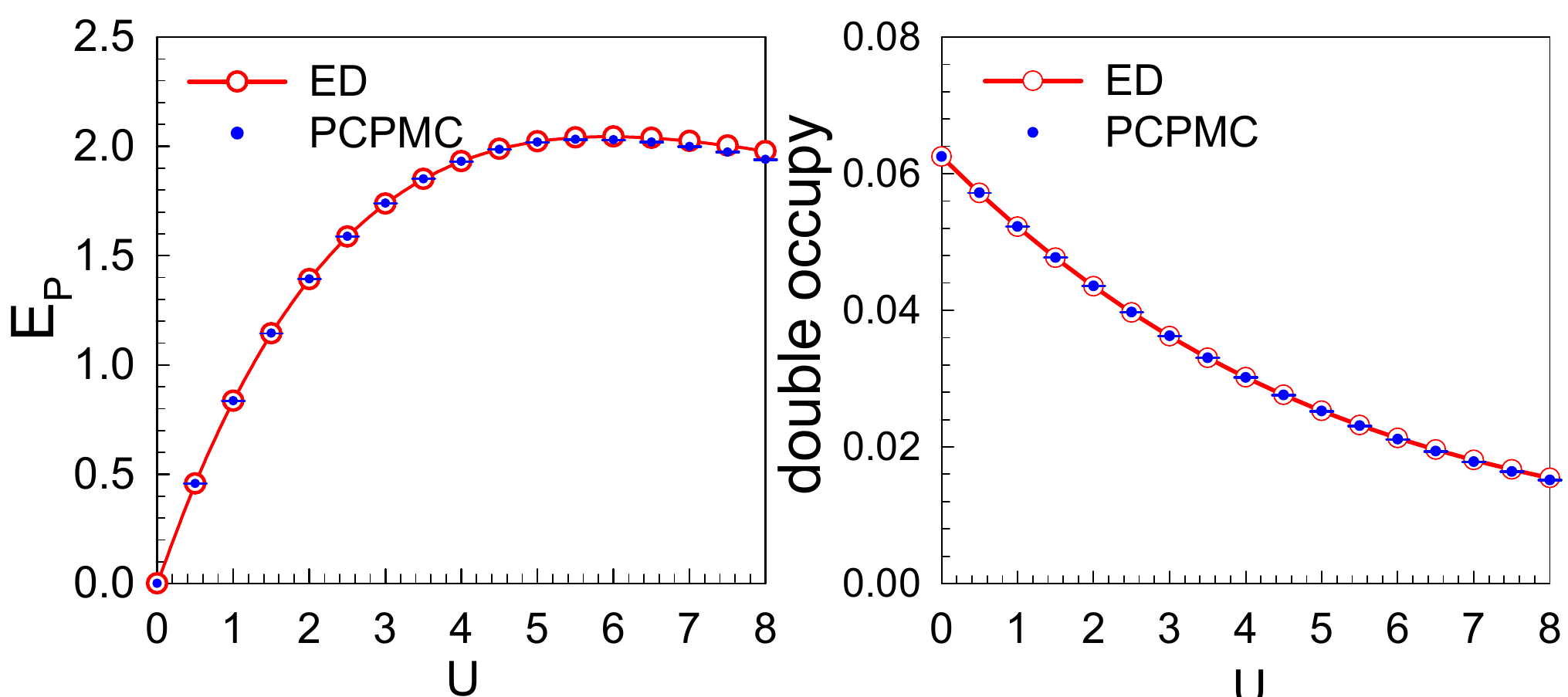}   
\caption{4$\times$4 Hubbard model in a magnetic field of 1/4 flux quanta per plaquette with 4 spin
up (and 4 down) electrons. (a) The potential energy and (b) the double occupy as a function of
$U$ for ED (red line with circles) and PCPMC (blue dots). The error bar is within the symbol.}
\label{Fig:ED}
\end{figure}

\section{Pairing correlation at open-shell fillings}

\begin{figure}[bp]
    \includegraphics[width=8cm]{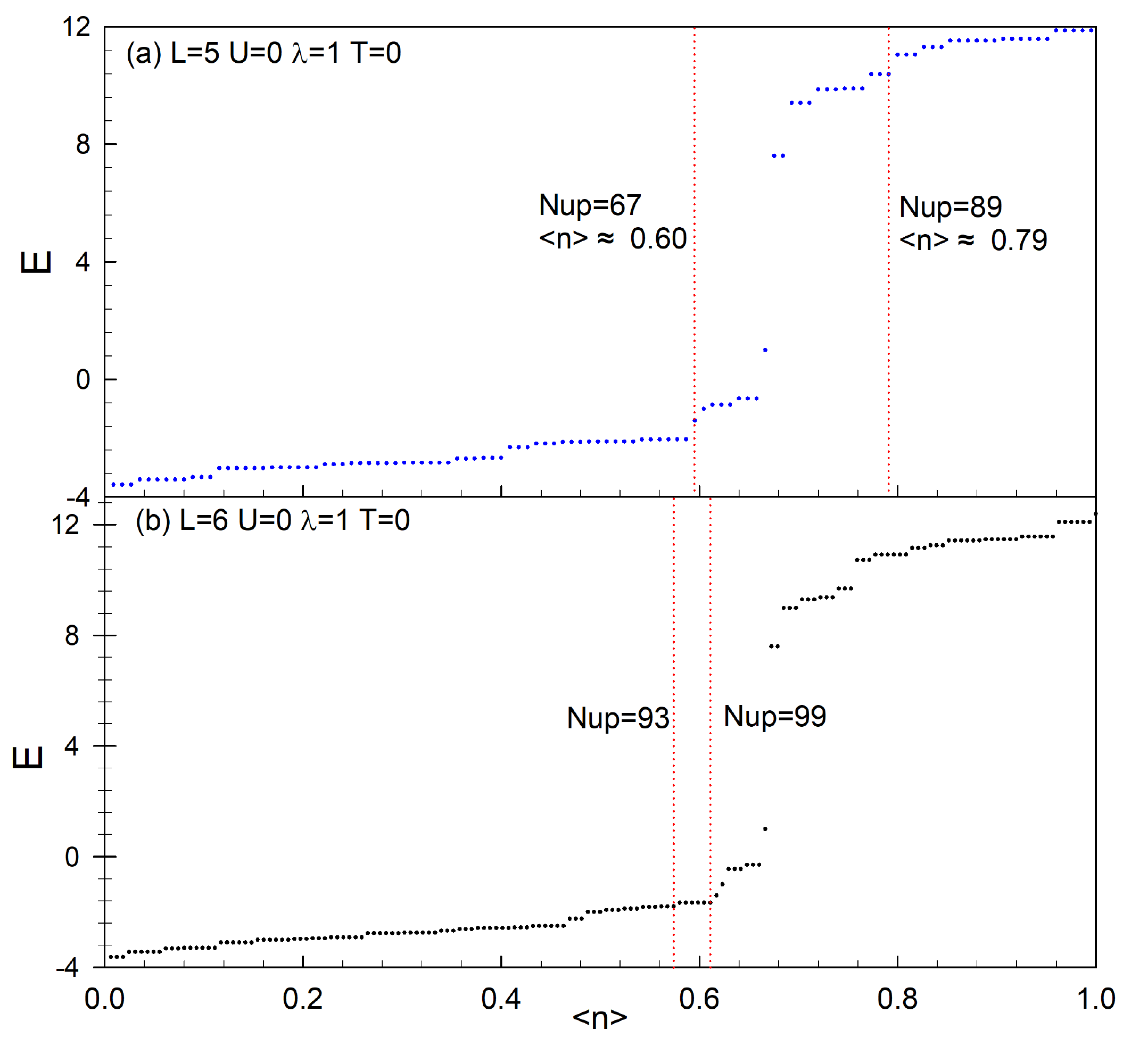}
    \caption{(Color online) Total energy $E$ as a function of electronic density $\avg{n}$ for the $L=5$ lattice with $\lambda=1$ at $T=0$ in the noninteracting limit ($U$=0).}
    \label{Fig:shell}
\end{figure}

We present more results on pairing correlation at open-shell fillings to further support the conclusion derived from close-shell fillings.
To have a global picture on the close-shell and open-shell fillings,
the total energy as a function of the electronic density $\avg{n}$ for a 3$\times3\times 5^2$ lattice (a) and 3$\times3\times 6^2$ lattice at $U=0$ has been shown in Fig.\ref{Fig:shell}.
Some close-shell fillings are marked by horizontal dashed lines in red color, and the open-shell fillings are on the ``platform", indicating the system degeneracy.

\begin{figure}[tbp]
\includegraphics[width=8.5cm]{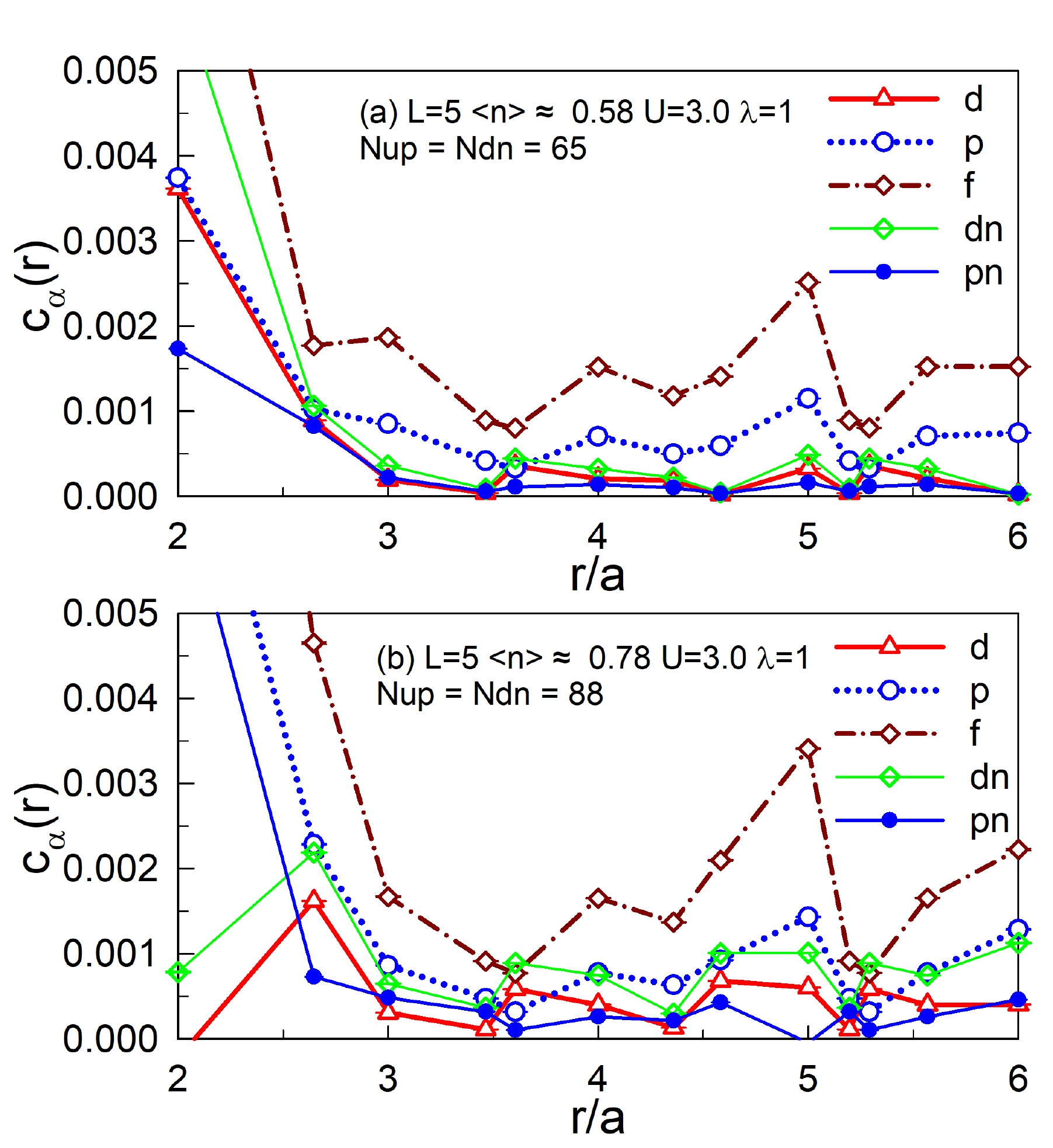}
\caption{The pairing correlation function for different pairing symmetries versus the distance $r$ at $\avg{n}\approx 0.58$ (a) and $\avg{n}\approx 0.78$ (b), which are both open-shell fillings. Here $U=3.0t$, $\lambda=1.0t$ and $L=5$.}
  \label{Fig:L5open}
\end{figure}

\begin{figure}[tbp]
\includegraphics[width=8.5cm]{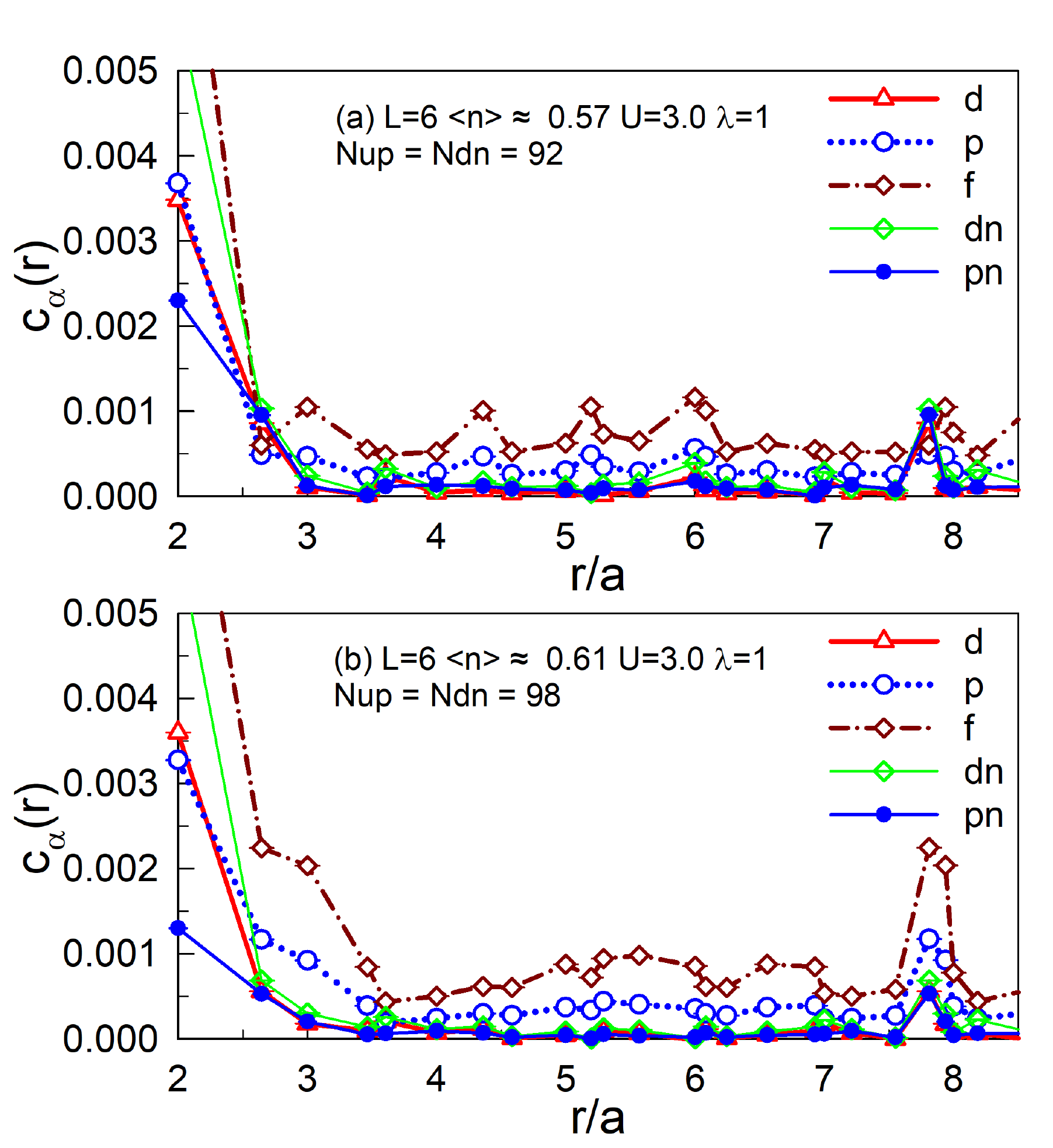}
\caption{The pairing correlation function for different pairing symmetries versus the distance $r$ at $\avg{n}\approx 0.57$ (a) and $\avg{n}\approx 0.61$ (b), which are both open-shell fillings. Here $U=3.0t$, $\lambda=1.0t$ and $L=6$.}
  \label{Fig:L6open}
\end{figure}

In Fig. \ref{Fig:L5open}, the pairing correlation with different symmetries are shown at $N_{up}=N_{dn}=65$, $\avg{n}\approx 0.58$ (a) and $N_{up}=N_{dn}=88$, $\avg{n}\approx 0.78$ (b), which are all at open-shell filling.
As they are at open-shell filling, we use the unrestricted Hartree-Fock wave functions as the trial wave function.
One can see that, within the filling range that we studied, the pairing correlation function with $f-$ wave symmetry dominate, which are consistent with that of close-shell fillings.

The similar results are shown for $L=6$ in Fig. \ref{Fig:L6open}, in which pairing correlation at two open-shell fillings, $N_{up}=N_{dn}=92$ (a) and $N_{up}=N_{dn}=98$ (b) are shown.
Again we see that, the pairing correlation with $f$-wave symmetry dominates over other symmetries.
These data at open-shell fillings further support the conclusion in the main body of this paper.

\bibliography{referenceMoS2}

\end{document}